\documentstyle[aps, pre, epsf]{revtex}
\begin{document}

\title{The origin of the logarithmic singularity in the symplectic semiclassical form factor}

\author{Stefan Heusler}
\address{heusler@theo-phys.uni-essen.de,
Fachbereich Physik, Universit\"at--Gesamthochschule Essen,
45\,117 Essen, Germany }
\date{\today}
\maketitle
\begin{abstract}
€

The semiclassical origin of the logarithmic singularity at the Heisenberg time of the symplectic form factor is deduced by combining the result of M. Sieber and K. Richter for the first term of the loop-expansion in the orthogonal case with the contribution that arises due to the spin.

We are able to make a quantitative statement about the topology of all non-diagonal contributions in terms of integrals over $SU(2)$ leading to the conclusion that the same perturbative loop expansion is responsible for the form factor in the region $0 < \tau < 2$ in the orthogonal and symplectic case taking into account Kramers' degeneracy; the only difference being a phase factor arising due to the spin. 

\end{abstract}€
€
\section{Introduction}

The analytic proof of the equivalence of random matrix theory with the spectral properties of classically chaotic quantum systems (BGS-conjecture) is still an open problem. Numerically, it was proven for many chaotic quantum systems that the form factor is faithful to random matrix theory \cite{ref0}, however, a quantitative explanation is still missing.

Recently, M. Sieber and K. Richter have been able to make a first important step towards a deeper understanding of this relation by deriving the first off-diagonal term of the semiclassical form factor for the orthogonal case \cite{ref1}. 

This paper is organized as follows. First, we will derive the first off-diagonal term of the symplectic form-factor using the semiclassical spin formalism derived by J. Bolte and S. Keppeler \cite{ref2}. Next, we are able to make a statement about all topological charges ($SU(2)$-Berry-phases) in the $\tau$-expansion of the form factor. This will lead to the final conclusion that the loop expansion for the orthogonal and symplectic case are basically the same in the region $0 < \tau < 2$; the only difference being the phase factor due to the spin. The main conclusion is that the singularity at the Heisenberg time in the symplectic case is originating from this additional phase factor.

The orthogonal form factor derived from random matrix theory is given by

\[
K[\tau]_{\rm orth.} = \left\{
\begin{array}{r@{\quad:\quad}l}
2 | \tau | - | \tau | \ln [1 + 2 | \tau | ] & | \tau | < 1
\\
2 - | \tau | \ln \frac{ 2 | \tau | + 1}{ 2 | \tau| - 1} & | \tau | > 1
\end{array} \right. \]

leading to a small $\tau$-expansion $K[\tau] = 2 \tau - 2 \tau^2 + ...$. M. Sieber demonstrated that the first off-diagonal contribution term $ - 2 \tau^2$ can be derived by considering the correlation of one self-intersecting with one non-self-intersecting loop,
that is, a pair of loops where the relative intersection number is one. These configurations do only contribute to the form factor when the time-reversal
symmetry is preserved. The reason is that the action difference of the two loops must be smaller than $\hbar$ in order to give a non-vanishing contribution to the form factor. Otherwise, the summation over all contributing loop pairs would have a random phase and vanishes. Taking an analogy from the theory of disordered systems, imagine a system being invariant under time reversal. By applying a magnetic field, the time reversal symmetry is destroyed. When the magnetic field is very weak, the classical paths of motion do not deviate strongly from the paths in the original time-reversal invariant system, however, when the magnetic flux through the surface of the two parts of the loop is larger than $\hbar$, the action difference leads to destructive interference, because although the enclosed area of the two loops is almost the same, the direction in one part of the first loop is reversed compared to the second loop. In the semiclassical limit, loops of length $L \rightarrow \infty$ have to be considered, therefore,
for a constant magnetic field, the action difference will be larger than $\hbar$ for the relevant loop pairs. Indeed, no $\tau^k$ terms, $k >1$, exists in the unitary case in the form factor. 

However, for the symplectic case, such off-diagonal terms arise much the same as for the orthogonal case. The symplectic form-factor derived from random matrix theory reads

€

\[ K[\tau]_{\rm sympl.} = \left\{
\begin{array}{r@{\quad:\quad}l}
\frac{| \tau |}{2} - \frac{|\tau|}{4} \ln | 1 - |\tau|| & | \tau | < 2
\\
1 \ \ \ \ \ \ \ \ \ \ \ \ \ \ \ \ \ \ \ \ \ \ & | \tau | > 2
\end{array} \right. \]

€

thus the term $ + \tau^2 /4 $ is expected to arise from the same loop correlation that has been 
considered by M. Sieber. The prefactor $1/4$ can be determined trivially due to Kramers' degeneracy: When the form factor $K( \tau )$ has been calculated using semiclassical analysis for the orthogonal case,
the change due to Kramers' degeneracy in the symplectic case leads to the relation \cite{ref2}

\begin{equation}
{\tilde K}(\tau) = \frac{1}{2} K(\frac{1}{2} \tau)
\end{equation}

not taking into account the spin contribution. In the symplectic case, the semiclassical first-order 
off-diagonal contribution reads therefore 

\begin{equation}
K_{\rm sympl.}^{\rm off} (\tau) = {\rm (spin \ contribution)} * \frac{1}{2} ( - 2 (\frac{\tau}{2})^2)
\end{equation}

Thus, the only non-trivial change for the symplectic case as compared to the orthogonal case is the
sign factor, which is, as we intend to show, due to the spin contribution in the semiclassical form factor. 

It was demonstrated that the spin contribution is $+1$ for the diagonal part of the form factor, that is, when both loops $\gamma$ are equal \cite{ref2}. However, the first off-diagonal contribution is due to intersecting loops, that is, non-trivial topology. 

\section{Trivial Topology-Case}

In general, the spin contribution to the density of states is given by \cite{ref2}

\begin{equation}
\rho (E)_{\rm osc} = \frac{1}{\pi \hbar} \sum_{r=1}^{\infty} 
\sum_{\alpha} {\rm Tr} \ d_{\gamma}^r \frac{T_{\alpha}}{\sqrt{ \det[M_{\alpha}^r - 1] } } 
\cos [ \frac{r}{\hbar} S_{\alpha} (E)] 
\end{equation}

Comparing with the usual formulation, a factor ${\rm Tr} \ d_{\gamma}^r$ arises due to the spin, 
where $d_{\gamma}$ is a matrix from $SU(2)$ obeying the differential equation

\begin{equation}
\label{eq3}
{\dot d}(p, x, t) + i M ( \Phi_H^t ( p, x) ) d(p, x, t) = 0
\end{equation}

Here, $\Phi_H^t ( p, x)$ describes the classical trajectory along a path $\gamma$, and 
$M$ is a hermitian and traceless matrix on the energy shell $\Omega_E$ in phase space describing the spin interaction. In the semiclassical limit, $t \rightarrow \infty$, it is assumed that the flow on the product space ${\cal M} := \Omega_E \times {\rm SU}(2)$ is strongly mixing, that is, the matrix $d(p, x, t) g$ has lost all memory about its initial conditions at $t=0$ and can be assumed to be an arbitrary ${\rm SU}(2)$ matrix $h$. In \cite{ref3}, quantum ergodicity of the
spin system has been proven under the assumption that the classical motion is ergodic. Basically, this is due to the Hopf map $\pi_H : SU(2) \rightarrow SO(3)$ relating the classical spin motion with the quantum spin motion. 

Using this assumption, $< {\rm Tr} \ d_{\gamma} {\rm Tr} \ d_{\gamma'} > = 1$ can be concluded for the diagonal contribution, that is, for $\gamma = \gamma'$.

First, we recall the key argument in the derivation of the spin contribution for the diagonal part, then we shall show how non-trivial topology alters the result. The object that has to be calculated in the diagonal part of the form factor is the integral of $( {\rm Tr} \ d_{\gamma} )^2 $ for large time $t$ over the product space ${\cal M} = \Omega_E \times {\rm SU}(2)$, that is,

\begin{eqnarray}
< ({\rm Tr} \ d_{\gamma})^2 >
& = & 
{\rm lim}_{t \rightarrow \infty} \int_{SU(2)} \int_{\Omega_E} ( {\rm Tr} \ d( p, x, t) g g^{-1} )^2 
d \mu_E (p, x) d\mu_H (g)
\end{eqnarray}
Here, $d \mu_E (p, x)$ is the Liouville measure on the energy shell $\Omega_E$, and $d\mu_H (g)$ is the Haar measure of ${\rm SU}(2)$.
Using the mixing condition $d( p, x, t) g \rightarrow h$ for large times $t$, this integral becomes in the limit $t \rightarrow \infty$
\begin{eqnarray}
< ({\rm Tr} \ d_{\gamma})^2 >
& = & 
\int_{SU(2)} \int_{SU(2)} ( {\rm Tr} \ h g^{-1} )^2 
d \mu_H(h) d\mu_H (g)
\end{eqnarray}
The Haar measure is right-invariant, therefore, the result is \cite{ref2}
\begin{equation}
< ({\rm Tr} \ d_{\gamma})^2 > = 
\int_{SU(2)} ( {\rm Tr} h )^2 d \mu_H(h) = +1
\end{equation}
Obviously, the spin contribution is irrelevant for the diagonal part of the 2-point function and the form factor.

From the transport equation (4) of $d(p, x, t)$, it follows the composition law

\begin{equation}
d(p, x, t + t') = d( \Phi_H^{t'} ( p, x), t) d(p, x, t')
\end{equation}

Therefore, it is possible to cut the loop at an arbitrary point (time $t'$)
into two pieces $d_{\gamma} = d_1 * d_2 = d_{\gamma'}$ and express the same integral as

\begin{eqnarray}
< ({\rm Tr} \ d_{\gamma})^2 >
& = & 
{\rm lim}_{t \rightarrow \infty} \int_{SU(2)} \int_{SU(2)} \int_{\Omega_E} ( {\rm Tr} \
[ d( \Phi_H^{t'} ( p, x), t) {\tilde g} {\tilde g}^{-1}  d( p, x, t') g g^{-1} ) ]^2 
d \mu_E (p, x) d\mu_H (g) d\mu_H ({\tilde g})
\end{eqnarray}
leading to 
\begin{eqnarray}
< ({\rm Tr} \ d_{\gamma})^2 >
& = & 
\int_{SU(2)} \int_{SU(2)} \int_{SU(2)} \int_{SU(2)} ( {\rm Tr} \ h g^{-1} {\tilde h} {\tilde g}^{-1})^2 
d \mu_H(h) d\mu_H (g) d \mu_H({\tilde h}) d\mu_H ({\tilde g}) = +1
\end{eqnarray}
Using the right-invariance of the Haar-measure, of course, the
same result is obtained. That is, cutting the two loops without reversing the
direction of one of the two loops does not alter the result, as it should be.
Using the symbol $S_{\gamma}$ for a part of the two loops having the same direction and $O_{\gamma}$ for a part with opposite direction, the
conclusion is $S_{1} S_{2} = S_{1+2}$ and $O_{1} O_{2} = O_{1+2}$.

In the next section, we consider the case of a loop with one intersection, that
is, a loop with the structure $S_{1} O_{2}$.

\section{Non-trivial Topology: One intersection-Case}

Next, we consider the general integral

€
\begin{eqnarray}
< {\rm Tr} \ d_{\gamma} {\rm Tr} \ d_{\gamma'}>
& = & 
{\rm lim}_{t \rightarrow \infty} \int_{SU(2)} \int_{\Omega_E} ( {\rm Tr} \ d_{\gamma} ( p, x, t) ) 
( {\rm Tr} \ d_{\gamma'} ( p, x, t) )
d \mu_E (p, x) d\mu_H (g)
\end{eqnarray}

€

For the first off-diagonal contribution, the direction of one part of one loop must be reversed, that is, cutting the first loop into two parts, $d_{\gamma} = d_1 * d_2$, the other loop is given by $d_{\gamma'} = \{ d_1 * d_2^{-1}, d_1^{-1} * d_2 \} $. These two possibilities correspond to $S_{1} O_{2}$ and $S_{2} O_{1}$, respectively. In the calcuation of the form factor, all possible
combinations of loops have to be taken into account. Therefore, the spin contribution is given by
$< {\rm Tr} \ d_{\gamma} {\rm Tr} \ d_{\gamma'}> = S_{1} O_{2} + S_{2} O_{1}$.

$$
< {\rm Tr} \ d_{\gamma} {\rm Tr} \ d_{\gamma'}>_{\rm 1  \ intersection} =
$$
$${\rm lim}_{t \rightarrow \infty} \int_{SU(2)} \int_{SU(2)} \int_{\Omega_E}  {\rm Tr} \  [  d_1 {\tilde g} {\tilde g}^{-1}  d_2 g g^{-1}  ]
\Big( 
{\rm Tr} \ [ d_1 {\tilde g} {\tilde g}^{-1}  g g^{-1} d_2^{-1} ] +
{\rm Tr} \
[  {\tilde g} {\tilde g}^{-1} d_1^{-1} d_2    g g^{-1} ] \Big)
d \mu_E (p, x) d\mu_H (g) d\mu_H ({\tilde g}) = 
$$
$$ 2 \int_{SU(2)} \int_{SU(2)} \int_{SU(2)} \int_{SU(2)}  
{\rm Tr} [ {\tilde h} {\tilde g}^{-1} h g^{-1} ]
{\rm Tr} [ {\tilde h} {\tilde g}^{-1} g h^{-1} ] 
 d\mu_H (h) d\mu_H (g)   d\mu_H ({\tilde h}) d\mu_H ({\tilde g}) = $$
$$2 \int_{SU(2)} \int_{SU(2)}
{\rm Tr} [ {\tilde h}  h  ]
{\rm Tr} [ {\tilde h} h^{-1} ]
 d\mu_H (h) d\mu_H ({\tilde h}) =$$
\begin{equation}
-2 \int_{-1}^{1} dx_1 \int_{-1}^{1} dx_2 
[\frac{1}{2} (1 - x_1)(1 - x_2)] \frac{1}{(8 \pi^2)^{2}} (2 \pi)^4=-1
\end{equation}

Thus, we have calculated the spin contribution for the first off-diagonal term for the symplectic case. The parameterization of the integral and more general expressions are given in the appendix. Combining this result with the results of M. Sieber and K. Richter and taking into account Kramers' degeneracy, the conclusion is

\begin{equation}
K(\tau)^{\rm off}_{\rm sympl.} =\frac{1}{2} [ - 2 (\frac{\tau}{2})^2 ]  < {\rm Tr} \ d_{\gamma} {\rm Tr} \ d_{\gamma'}>_{\rm 1  \ intersection} = + \frac{\tau^2}{4}
\end{equation}
as presumed.

\section{Non-trivial Topology: Multiple intersection-Case}

Next, we want to derive a general statement about the topology of higher
order terms in the $\tau$-expansion. By comparing the orthogonal and the symplectic form factor for higher orders in $\tau$ and anticipating random-matrix behavior, it is possible to predict the $SU(2)$-topological charge and thus the contributing topology for all higher powers of $\tau$: Assume that semiclassically, the orthogonal form
factor $K(\tau)$ has been evaluated, that is, the expansion in $\tau$ has been deduced by calculating the correlation between non-trivial topology of different loops, then $1/2 \ K( \tau /2)$ corresponds to the symplectic form factor without taking into account the spin contribution. Indeed, comparing the power expansion of $K_{\rm sympl} (\tau)$ with $\frac{1}{2} \ K_{\rm orth} (\frac{\tau}{2})$, the result is

€

\begin{equation}K_{\rm sympl}(\tau) = \frac{\tau^2}{4} + \frac{\tau^3}{8} + \frac{\tau^4}{12} + \frac{\tau^5}{16} + ...
\end{equation}

\begin{equation}
1/2 \ K_{\rm orth}(\tau /2) = - \frac{\tau^2}{4} + \frac{\tau^3}{8} - \frac{\tau^4}{12} + \frac{\tau^5}{16} - ...
\end{equation}

€

Thus, the conclusion for the spin contribution (the topological charge) for all off-diagonal contributions is

\begin{equation}
< {\rm Tr} \ d_{\gamma} {\rm Tr} \ d_{\gamma'}>_{\rm k \ intersections} = (-1)^{k}
\end{equation}

This equation should be interpreted as follows: Taking the loops $\gamma, \gamma'$ corresponding to
the $\tau^{k+1}$-term of the expansion, the topological charge due to the spin must be $(-1)^{k}$ if the BGS-conjecture is
correct. The term $k=0$ has been calculated in \cite{ref2}, $k=1$ has been calculated in the previous section, corresponding to the configuration with relative intersection number $k=1$). 

Next, we consider the case where the relative intersection number is $k=2$. For every new
intersection, a combination same direction - opposite direction $S_{\alpha} O_{\beta}$ must be added to the existing loop. Obviously, summation over all possible spin contributions for the loops leads to $2^k$ terms, which all give the same contribution in the semiclassical limit $t \rightarrow \infty$.

$$< {\rm Tr} \ d_{\gamma} {\rm Tr} \ d_{\gamma'}>_{\rm 2  \ intersections} =$$
$$2^2 \int_{SU(2)} \int_{SU(2)} \int_{SU(2)} \int_{SU(2)}  
{\rm Tr} [ h_1 h_2 h_3 h_4 ]  {\rm Tr} [ h_1 h_2^{-1} h_3 h_4^{-1}]  
 d\mu_H (h_1) d\mu_H (h_2) d\mu_H (h_3) d\mu_H (h_4) =$$
\begin{equation}
2^2 \int_{-1}^{1} dx_1 \int_{-1}^{1} dx_2 \int_{-1}^{1} dx_3 \int_{-1}^{1} dx_4 
[\frac{1}{4} (1 - x_1)(1 - x_2)(1 - x_3)(1 - x_4)] \frac{1}{(8 \pi^2)^{4}} (2 \pi)^8=+1
\end{equation}
We conclude that the $\tau^3$-term in the expansion of
the form factor both in the orthogonal and symplectic case is due to correlations between orbits where the relative intersection number is two.

Next, we shall give the general expression for the spin contribution in the
form factor when the relative intersection number is $k$. The prefactor $2^k$ is due to the
symmetrization of the expression in each of the $k$ segments $\prod_{j=1}^{k} S_{\alpha_j} O_{\beta_j}$ of the loop.

$$< {\rm Tr} \ d_{\gamma} {\rm Tr} \ d_{\gamma'}>_{\rm k  \ intersections} =$$
$$ 2 ^{k} 
\prod_{j = 1}^{2 k}\int_{SU(2)_j}  d\mu_H (h_j) \ 
{\rm Tr} [\prod_{l=1}^{2 k} h_l ] 
 \ {\rm Tr} [\prod_{t =1}^{k} h_{2 t - 1} h_{2 t}^{-1}] 
=$$
\begin{equation}
(-1)^k \prod_{j=1}^{2 k} \int_{-1}^{1} d x_j (1 - x_j) \frac{1}{(8 \pi^2)^{2k}} (2 \pi)^{4 k} = (-1)^k 
\end{equation}
as presumed.

Finally, we compare $K[\tau]_{\rm sympl.}$ with $\frac{1}{2} K[\tau /2]_{\rm orth.}$ and find

\[
\frac{1}{2} K[\frac{\tau}{2} ]_{\rm orth.} = \left\{
\begin{array}{r@{\quad:\quad}l}
 \frac{| \tau |}{2} - \frac{| \tau  |}{4} \ln [1 + | \tau | ] & | \tau | < 2
\\
1 - \frac{| \tau |}{4} \ln[\frac{1 + | \tau |}{1 - | \tau|}] & | \tau | > 2
\end{array} \right. \]

\[
K[\tau]_{\rm sympl.} = \left\{
\begin{array}{r@{\quad:\quad}l}
 \frac{| \tau |}{2}  - \frac{| \tau |}{4} \ln [1 - | \tau | ] & | \tau | < 2
\\
1 \ \ \ \ \ \ \ \ \ \ \ \ \ \ \ \ \  & | \tau | > 2
\end{array} \right. \]

Obviously, the sign factor which has been demonstrated to be related to the spin contribution
is responsible for the singularity at $\tau = 1$ in the symplectic case. 

To summarize, we have been able to show that the BGS-conjecture leads to
the conjecture that the term proportional to $\tau^{k+1}$ in the small $\tau$-expansion both of the symplectic and orthogonal form factor is generated by correlations of loops having the relative intersection number $k$. Using this conjecture, we have demonstrated that the semiclassical loop expansion for the off-diagonal part of the form factor in terms of correlations between the loops $\gamma, \gamma'$, with relative intersection number $k$ reproduces correctly the singularity at the Heisenberg time arising in the symplectic case.

I am grateful to P. Braun, M. Sieber, S. Keppeler, J. Bolte, C. Manderfeld, H. Schomerus and F. Haake for useful discussions.

\newpage

\section{Appendix}

In this appendix we give the parameterization and the most important steps
for the evaluation of the integrals over the group $SU(2)$.
A general $SU(2)$-matrix can be expressed as ($q \equiv \exp( i \eta), r \equiv \exp ( i \xi)$)

\[
h(q, r, \theta)  =   \left(
        \begin{array}{l@{\quad \quad}r}
                                   \cos(\theta) \ q & \sin(-\theta) \ r \\ \sin(\theta) \ r^{-1}  & \cos(\theta) \ q^{-1}
        \end{array}     
       \right)
\]

The integration measure in terms of the variables $(\theta, \xi, \eta)$ reads

$$ 
\int_{SU(2)} d \mu_H (h) = \int_{0}^{ \pi/2}  \int_{0}^{ 2 \pi} 
\int_{0}^{ 2 \pi} \frac{1}{4 \pi^2} \sin(2 \theta) d\theta \  d\xi  \ d\eta
$$

Now, consider a general integral of the form

$$< {\rm Tr} \ d_{\gamma} {\rm Tr} \ d_{\gamma'}>_{\rm k  \ intersections} =$$
$$ 2 ^{k} 
\prod_{j = 1}^{2 k}\int_{SU(2)_j}  d\mu_H (h_j)
{\rm Tr} [\prod_{l=1}^{2 k} h_l ] \ 
{\rm Tr} [\prod_{t =1}^{k} h_{2 t - 1} h_{2 t}^{-1}] 
=
 2 ^{k} 
\prod_{j = 1}^{2 k}\int_{SU(2)_j}  d\mu_H (h_j) 
f(q_j, r_j, \theta_j)
$$

Written in terms of $(q_j, r_j)$, the function $f(q_j, r_j, \theta_j)$ is a polynomial in  $\prod_{j= 1}^{2 k} q_j^{n_j} r_j^{m_j}$. The $4k$ integrals over $(\eta, \xi)$ extract the term proportional to $\prod_{j= 1}^{2 k} q_j^0 r_j^0 =1$ of $f(q_j, r_j, \theta_j)$, which will be denoted as $g(\theta_j)$. The remaining $2k$ integrals are then given by

$$< {\rm Tr} \ d_{\gamma} {\rm Tr} \ d_{\gamma'}>_{\rm k  \ intersections} =
 2 ^{k} 
\prod_{j = 1}^{2 k}\int_{0}^{\pi/2} d \cos(2 \theta_j) (\frac{1}{8 \pi^2})^{2k} 
(2 \pi)^{4 k} g(\sin(\theta_j)^2, \cos(\theta_j)^2)=
$$
$$
 2 ^{k} 
\prod_{j = 1}^{2 k}\int_{-1}^{1}  d x_j  (\frac{1}{8 \pi^2})^{2k} (2 \pi)^{4 k}
g((1 - x_j )/2, (1 + x_j)/2)=
(-1)^k \prod_{j=1}^{2 k} \int_{-1}^{1} d x_j (1 - x_j) \frac{1}{(8 \pi^2)^{2k}} (2 \pi)^{4 k} = (-1)^k $$

The functions $f(q_j, r_j, \theta_j)$ and $g(\theta_j)$ have been determined with the help of Mathematica.

\end{document}